\documentclass[12pt]{article}
\usepackage{epsfig}
\newcommand{\be}{\begin{eqnarray}}
\newcommand{\ee}{\end{eqnarray}}
\newcommand{\MeV}{{\rm\,MeV}}
\newcommand{\GeV}{{\rm\,GeV}}

\newcommand{\slk}{/\kern-6pt k}
\newcommand{\slp}{p\kern-5pt/}

\newcommand{\pfrac}[2]{\left(\frac{#1}{#2}\right)}

\begin{document}
\thispagestyle{empty}
\begin{flushright}
UM-TH-00-30\\
MZ-TH/00-33\\
CLNS-00/1686\\
hep-ph/0008156\\
\end{flushright}
\vspace{0.5cm}
\begin{center} 
{\Large\bf Top quark mass definition and}\\[3mm]
{\Large\bf $t\bar t$ production near threshold at the NLC}\\[1.7cm]
{\sc\bf Oleg Yakovlev$^{1,a}$ and Stefan Groote$^{2,3,b}$}\\[1cm] 
$^1$ Randall Laboratory of Physics, University of Michigan,\\[2mm]
  Ann Arbor, Michigan 48109-1120, USA\\[5mm]
$^2$ Institut f\"ur Physik der Johannes-Gutenberg-Universit\"at,\\[2mm]
  Staudinger Weg 7, 55099 Mainz, Germany\\[5mm]
$^3$ Floyd R. Newman Laboratory, Cornell University,\\[2mm]
  Ithaca, NY 14853, USA\\
\end{center}
\vspace{2cm}

\begin{abstract}\noindent
We suggest an infrared-insensitive quark mass, defined by subtracting the soft
part of the quark self energy from the pole mass. We demonstrate the deep
relation of this definition with the static quark-antiquark potential. At
leading order in $1/m$ this mass coincides with the PS mass which is defined
in a completely different manner. Going beyond static limit, the small
normalization point introduces recoil corrections which are calculated here as
well. Using this mass concept and other concepts for the quark mass we
calculate the cross section of $e^+e^-\to t\bar t$ near threshold at NNLO
accuracy adopting three alternative approaches, namely (1) fixing the pole
mass, (2) fixing the PS mass, and (3) fixing the new mass which we call the
$\overline{\rm PS}$ mass. We demonstrate that perturbative predictions for the
cross section become much more stable if we use the PS or the
$\overline{\rm PS}$ mass for the calculations. A careful analysis suggests
that the top quark mass can be extracted from a threshold scan at NLC with an
accuracy of about $100-200\MeV$.
\end{abstract}
\vspace*{\fill}

\noindent $^a${\small e-mail: yakovlev@umich.edu}\\
\noindent $^b${\small e-mail: groote@thep.physik.uni-mainz.de}

\newpage
\section{Introduction}
One of the main goals of future $e^+e^-$ and $\mu^+\mu^-$ colliders such as
the Next Linear Collider (NLC) and the Future Muon Collider (FMC) will be to
measure and to determine the properties of the top quark which was first
discovered at the Tevatron~\cite{Fermi} with a mass of
$m=174.3\pm 5\GeV$~\cite{Fermi1}. Although the top quark will be studied at
the LHC and the Tevatron (RUN-II) with an expected accuracy for the mass of
$2-3\GeV$, the {\em most accurate measurement} of the mass with an accuracy
of $0.1\%$ ($100-200\MeV$) is expected to be obtained only at the 
NLC~\cite{Peskin}. 

Due to the large top quark width, the top-antitop pair cannot hadronize into
toponium resonances. The cross section appears therefore to have a smooth
line-shape showing only a moderate $1S$ peak. In addition the top quark width
serves as an infrared cutoff~\cite{Fadin:1987wz,Fadin:1988fn} and as a natural
smearing over the energy~\cite{Poggio:1976af}. As a result, the nonperturbative
QCD effects induced by the gluon condensate are
small~\cite{Strassler:1991nw,Fadin:1991jh}, allowing us to calculate the cross
section with high accuracy by using perturbative QCD even in the threshold
region. Many theoretical studies at LO and NLO have been done in the past for
the total cross
section~\cite{Fadin:1987wz,Fadin:1988fn,Strassler:1991nw,Kwong:1991iy}, 
for the momentum distribution~\cite{Jezabek:1992np,Sumino:1993ai}, also
accounting for electro-weak corrections~\cite{Guth, Hollik,FadYak1}, and for
the complete NLO correction including non-factorizable
corrections~\cite{Melnikov:1994np,Fadin:1994dz,Fadin:1994kt,SuminoTH}.

Recently, the NNLO analysis of the inclusive threshold production cross
section has been reported~\cite{Hoang:1998xf,Melnikov:1998pr,Yakovlev:1999ke,
  Beneke:1999qg,Hoang:1999zc,Nagano,Penin}. The results of the NNLO analysis
are summarized in a review article~\cite{Review}. To summarize the results for
a standard approach using the pole mass, the NNLO corrections are
uncomfortably large, spoiling the possibility for the top quark mass
extraction at NLC with good accuracy because the $1S$ peak is shifted by about
$0.5\GeV$ by the NNLO, the last known correction. One of the main reasons for
this is the usage of the pole mass in the calculations. It was realized that
such type of instability is caused by the fact that the pole mass is a badly
defined object within full QCD~\cite{Beneke:1994sw,Vainshtein}.

In this paper we suggest a definition of the quark mass alternative to the
pole mass. We call it $\overline{\rm PS}$ mass because in the static limit
this mass coincides with the potential subtracted (PS) mass, even though the
PS mass is defined in a different manner~\cite{Beneke}. This is the reason why
we include both of them in this paper. The ratio of the small normalization
point and the mass introduce relativistic corrections (called recoil
corrections) which result in the difference between the two masses at higher
orders in $1/m$. In contrast to the pole mass, the PS and the
$\overline{\rm PS}$ mass are not sensitive to non-perturbative QCD effects. We
derive recoil corrections to the relation of the pole mass to the
$\overline{\rm PS}$ mass and demonstrate that perturbative predictions for the
cross section become much more stable at higher orders of QCD (shifts are
below $0.1\GeV$) if we use the PS or $\overline{\rm PS}$ mass for the
calculations, as it is the case for any of the other threshold masses. This
understanding removes one of the obstacles for the accurate top quark mass
measurement and it can be expected that the top quark mass will be extracted
from a threshold scan at NLC with an accuracy of about $100-200\MeV$. The
necessity of isolating the IR contributions in the mass calculation and the
consideration of a short distance mass have been studied
intensively~\cite{Vainshtein,Beneke,Uraltsev1,Uraltsev2,Hoang1}. The 
applications of the PS, LS and $1S$ mass have been reported recently by
several groups~\cite{Beneke:1999qg,Hoang:1999zc,Nagano,Karlsruhe,MelnikovK}
and reviewed in Ref.~\cite{Review}. The detailed comparison of our results
using the PS mass with results of other groups have been performed in
Ref.~\cite{Review} so that we refer the reader to this reference for details.

This paper is therefore devoted to two subjects. First, we define the
$\overline{\rm PS}$ mass which allows us to calculate recoil corrections near
the threshold. Second, we calculate the cross section for $e^+e^-\to t\bar t$
near threshold with NNLO accuracy using three alternative mass concepts, i.e.\
(1) the pole mass, (2) the static PS mass, and (3) the new $\overline{\rm PS}$
mass. The results of the first approach  has already been reported in
Ref.~\cite{Yakovlev:1999ke}, the {\em preliminary\/} results for the top quark
pair production near threshold using the static PS mass were presented at the
workshop ``Physics at Linear Colliders''~\cite{Karlsruhe,Review} by one of the
authors. The results on the second and third alternative concepts which we
present here are new.

The paper is organized as follows: In Sec.~2 we give the definition of the
$\overline{\rm PS}$ mass in terms of the soft part of the self energy
and calculate the leading order contribution. In Sec.~3 we calculate the
two-loop corrections and present our final result. In Sec.~4 we use the
obtained results for the analysis of the top quark pair production. In Sec.~5
we give our conclusions. The Appendix contains explicit expressions for the
coefficients which occur in the text.

\section{Mass definitions}
The top quark mass is an input parameter of the Standard Model. Although it is
widely accepted that the quark masses are generated due to the Higgs mechanism,
the value of the mass cannot be calculated from the Standard Model. Instead,
quark masses have to be determined from the comparison of theoretical
predictions and experimental data. 

It is important to stress that there is no unique definition of the quark mass.
Because the quark cannot be observed as a free particle like the electron,
the quark mass is a purely theoretical notion and depends on the concept
adopted for its definition. The best known definitions are the pole mass and
the $\overline {\rm MS}$ mass. However, both definitions are not adequate for 
the analysis of top quark production near threshold. The pole mass should not
be used because it has the renormalon ambiguity and cannot be determined more
accurately than $300-400\MeV$~\cite{Beneke:1994sw,Vainshtein} (see also
Refs.~\cite{Beneke,Scot}). The $\overline{\rm MS}$ mass is an Euclidean
 mass, defined at high virtuality, and therefore destroys the non-relativistic
expansion. Instead, it was recently suggested to use threshold masses like the
low scale (LS) mass~\cite{Vainshtein}, the potential subtracted (PS)
mass~\cite{Beneke}, or one half of the perturbative mass of a fictious
$1^3S_1$ ground state (called $1S$ mass) \cite{Hoang:1999zc}. In this paper we
study the static PS mass suggested in Ref.~\cite{Beneke},
\begin{eqnarray}\label{defPS}
m_{\rm PS}=m_{\rm pole}-\delta m_{\rm PS}\quad\mbox{with}\quad
\delta m_{\rm PS}=-\frac12\int^{|\vec k|<\mu_f}\frac{d^3k}{(2\pi)^3}
  V_C(|\vec k|)
\end{eqnarray}
where $V_C$ is the quark-antiquark Coulomb potential. In order to understand
why this mass definition is more adequate than the pole mass and to see that
the pole mass is very sensitive to long distance effects, it is enough to
consider the one-loop expression for the self energy diagram. Taking the
residue in $k_0$, one obtains a soft self energy contribution which comes from
momenta $k$ with $|\vec k|<\Lambda_{\rm QCD}$,
\begin{equation}\label{critical}
\delta m=4\pi\alpha_sC_F\kern-12pt
  \int\limits_{|\vec k|<\Lambda_{\rm QCD}}\kern-12pt
  \frac{d^3k}{(2\pi)^3}\frac1{2|\vec k|^2}
  =\frac{\alpha_s C_F}{\pi}\Lambda_{\rm QCD}.
\end{equation}
We observe that the pole mass has a non-perturbative uncertainty of order
$\Lambda_{\rm QCD}$ which then penetrates into consequent perturbative 
QCD calculations. It is easy to realize that the PS mass is free of this
ambiguity. Indeed, the term $\delta m$ in Eq.~(\ref{critical}) cancels in the
definition of the PS mass as given in Eq.~(\ref{defPS}) as well as in the
combination  $2m_{\rm pole}+V(r)$.  The definition in Eq.~(\ref{defPS}) has
been given in Ref.~\cite{Beneke} but has already been discussed implicitly in
Ref.~\cite{Vainshtein}. The remarkable step made in Ref.~\cite{Beneke} is to
use this definition beyond one-loop order. It has been proven in
Ref.~\cite{Beneke} that the cancellation of the infrared QCD contributions 
to the PS mass in  Eq.~(\ref{defPS}) holds even at higher loop orders.

In this paper we suggest a natural definition of a short distance quark mass
near threshold. Our objective is that the definition should be gauge
independent and well-defined within quantum field theory so that radiative and
relativistic corrections can be calculated in a systematic way. Nevertheless,
we stress that both concepts, as well as the other concepts mentioned before,
are equivalently applicable to considerations near the quark production
threshold.

\subsection{The $\overline{\rm PS}$ mass}
\begin{figure}
\begin{center}
\epsfig{figure=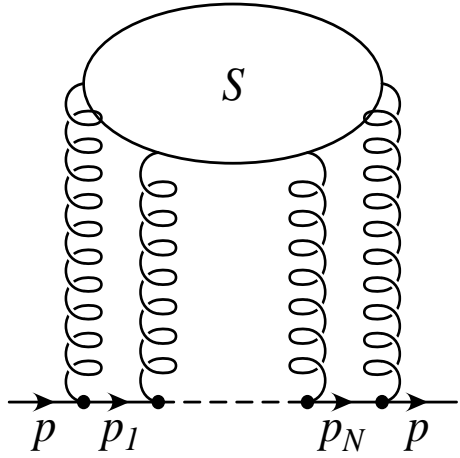, height=6truecm,width=6truecm}
\caption{\label{fig1}
  The general structure of the self energy diagram of a quark} 
\end{center}
\end{figure}

We define 
\begin{eqnarray}
m_{\overline{\rm PS}}=m_{\rm pole}-\delta m_{\overline{\rm PS}}
  \quad\mbox{with}\quad
  \delta m_{\overline{\rm PS}}=\Sigma_{\rm soft}(\slp)\Big|_{\slp=m}
\end{eqnarray}
where $\Sigma_{\rm soft}$ is the soft part of the heavy quark self
energy which is defined as the part where at least one of the heavy quark
propagators is on-shell.  A more precise definition is given below. We will
show that the static PS mass and the new $\overline{\rm PS}$ mass definition
coincide at leading order in $1/m$. At the same time we stress that the new
definition accounts for recoil corrections of orders $1/m$ and $1/m^2$. 
But first we describe the physical picture and discuss the structure of the
quark self energy shown in Fig.~\ref{fig1}. 

The starting point of our considerations is an on-shell quark with mass $m$ and
momentum $p$ (i.e.\ $p^2=m^2$) which we consider to be at rest,
$p=(m,\vec 0)$. This quark interacts with a number of gluons, the subdiagram
$S$ displayed in Fig.~\ref{fig1} describes the interaction between the gluons. 
In general the quark lines between the interaction points represent virtual
quark states. However, if the virtual quark comes very close to the mass shell
and the total momentum of the cloud of virtual gluons becomes soft, this
situation gives rise to long-distance nonperturbative QCD interactions. The
described (virtual) contributions result in the soft part  
of the self energy, $\Sigma_{\rm soft}$. 
For a precise definition we start with a general self 
energy diagram as shown in Fig.~\ref{fig1},
\begin{equation}\label{def1}
-i\Sigma(\slp)=\int\prod_{m=1}^M\frac{d^4l_m}{(2\pi)^4}
  S^{\{a_n\}}_{\{\alpha_n\}}(\{l_m\})
  \left(-ig_s\gamma^{\alpha_{N+1}}T_{a_{N+1}}\right)
  \prod_{n=N}^1\frac{i}{\slp_n-m}\left(-ig_s\gamma^{\alpha_n}T_{a_n}\right)
\end{equation}
where the last factor is a non-commutative product with decreasing index $n$.
The line momenta $k_n$ are linear combinations of the gluon loop momenta $l_m$,
the particular representation is specified by the structure $S$. The symbol
$\{l_m\}$ means the set of all these loop momenta, the same symbol is used
for the Lorentz and color indices. In general we have $M\le N$ which means
that line momenta can be correlated. The momenta of the virtual quark states
are given by $p_n=p+k_n$. Taking this as the starting point we define
\begin{eqnarray}\label{def2}
-i\Sigma_{\rm soft}(\slp)&=&\sum_{i=1}^N\int\prod_{m=1}^M
  \frac{d^4l_m}{(2\pi)^4}S^{\{a_n\}}_{\{\alpha_n\}}(\{l_m\})
  \left(-ig_s\gamma^{\alpha_{N+1}}T_{a_{N+1}}\right)\prod_{n=N}^{i+1}
  \frac{i}{\slp_n-m}\left(-ig_s\gamma^{\alpha_n}T_{a_n}\right)\
  \times\nonumber\\&&\times\ i(\slp_i+m)\left(-i\pi\delta(p_i^2-m^2)\right)
  \left(-ig_s\gamma^{\alpha_i}T_{a_i}\right)\prod_{n=i-1}^1
  \frac{i}{\slp_n-m}\left(-ig_s\gamma^{\alpha_n}T_{a_n}\right).
\end{eqnarray}
This equation is the definition of the soft part of the quark self energy. One
can derive this expression from Eq.~(\ref{def1}) by using the identity
\begin{equation}\label{ident1}
\frac1{p^2-m^2+i\epsilon}=-i\pi\delta(p^2-m^2)+P\pfrac1{p^2-m^2}
\end{equation}
and the fact that the principal value integral does not give any infrared
sensitive contribution. The delta function can be used to remove the
integration over the zero component of $k_i$. In order to parameterize the
softness of the gluon cloud we impose a cutoff on the spatial component,
$|\vec k_i|<\mu_f$, and indicate this by a label $\mu_f$ written at the upper
limit of the three-dimensional integral. This cutoff $\mu_f$ is also known as
{\em factorization scale\/}. So we can rewrite Eq.~(\ref{def2}) as
\begin{equation}
\Sigma_{\rm soft}(\slp,\mu_f)=-\frac12\sum_{i=1}^N\int^{\mu_f}
  \frac{d^3k_i}{(2\pi)^3}V(\vec k_i,p)
\end{equation} 
where
\begin{eqnarray}
V(\vec k_i,p)&:=&-\int\prod_{m=1}^{M-1}\frac{d^4l_m}{(2\pi)^4}
  S^{\{a_n\}}_{\{\alpha_n\}}(\{l_m\})
  \left(-ig_s\gamma^{\alpha_{N+1}}T_{a_{N+1}}\right)\prod_{n=N}^{i+1}
  \frac{i}{\slp_n-m}\left(-ig_s\gamma^{\alpha_n}T_{a_n}\right)\
  \times\nonumber\\&&\qquad\times\ \frac{\slp_i+m}{2p_i^0}
  \left(-ig_s\gamma^{\alpha_i}T_{a_i}\right)\prod_{n=i-1}^1
  \frac{i}{\slp_n-m}\left(-ig_s\gamma^{\alpha_n}T_{a_n}\right).
\end{eqnarray}
The range of the index $m$ is reduced by one which indicates that one of the
loop momenta is extracted as line momentum of the $i$-th line. In the following
we deal with the different realizations of this compact expression. As we will
see explicitly, the function $V(\vec k,p)$ occurring as integrand can be seen
as quark-antiquark potential where we have summed over the spin of the tensor
product of a final state with an initial state. Because the static
quark-antiquark potential is used in a similar way in Ref.~\cite{Beneke}, we
recover the result of Ref.~\cite{Beneke} in the static limit. However, there
is no kind of hierarchical order between both concepts because both solve the
afore mentioned problems with non-perturbative uncertainties of the order
$O(\Lambda_{\rm QCD})$. The role of non-perturbative effects of the order
$1/m$ have to be studied elsewhere.

\subsection{The leading order perturbative contribution}
The leading order contribution to the self energy of the quark is given by
\begin{equation}
\Sigma(\slp)=i\int\frac{d^4k}{(2\pi)^4}(-ig_s\gamma_\alpha T_a)
  \frac{i}{\slp+\slk-m}(-ig_s\gamma^\alpha T_a)\frac{-i}{k^2}
\end{equation}
where Feynman gauge is used for the gluon. The soft contribution thus reads
\begin{eqnarray}
\Sigma_{\rm soft}(\slp)&=&-ig_s^2C_F\int\frac{d^4k}{(2\pi)^4k^2}
  \gamma_\alpha(\slp+\slk+m)\gamma^\alpha
  \left(-i\pi\delta\left((p+k)^2-m^2\right)\right)\ =\nonumber\\
  &=&-\pi g_s^2C_F\int\frac{d^4k}{(2\pi)^4k^2}\left(-2(\slp+\slk)+4m\right)
  \delta\left((p+k)^2-m^2\right).
\end{eqnarray}
The projector $(1+\gamma^0)/2$ which represents the on-shell quark at rest can
be placed at the left and at the right of the Dirac structure and will lead to
a further simplification. In terms of the zero component of the momentum $k$
the Dirac delta function has two zeros $k_0=k_+$ and $k_0=k_-$ with
\begin{equation}
k_\pm:=\pm\sqrt{\kappa^2+m^2}-m
\end{equation}
where $\kappa=|\vec k|$. The delta function is therefore written as
\begin{equation}
\delta\left((p+k)^2-m^2\right)=\frac1{2\sqrt{\kappa^2+m^2}}
  \left(\delta(k_0-k_+)+\delta(k_0-k_-)\right).
\end{equation}
\begin{figure}
\begin{center}
\epsfig{figure=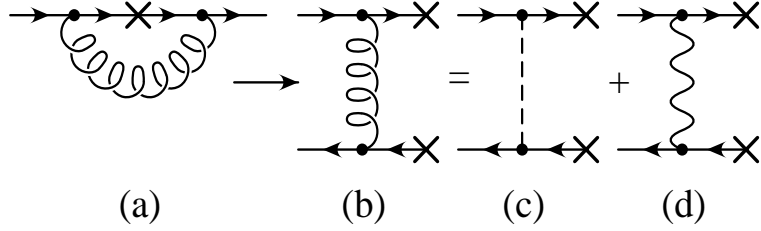, height=3truecm,width=10truecm}
\caption{\label{fig2}Leading order contribution to the quark self energy
(a) and to the quark-antiquark potential (b). The cross indicates the
point where we cut the quark line by imposing an on-shell condition to
the virtual quark state. The gluon propagator can be decomposed in a
Coulomb propagator (c) and a transverse propagator (d).} 
\end{center}
\end{figure}
The procedure which is done here is shown in Fig.~\ref{fig2}(a--b). The cross
indicates that we cut the line at this point by imposing the on-shell
condition to the corresponding (virtual) momentum. The diagram then proceeds
to a quark-antiquark interaction diagram where we have kept the crosses to
indicate the position of the cut line. This line carries the momentum $p+k$
while the other two external lines carry the momentum $p$. 
 Accordingly we obtain
\begin{equation}\label{leadsoft}
\Sigma_{\rm soft}(\slp,\mu_f)=-\frac12\int^{\mu_f}\frac{d^3k}{(2\pi)^3}
  V(\vec k,p),\qquad V(\vec k,p)=V_+(\vec k,p)+V_-(\vec k,p)
\end{equation}
where
\begin{equation}\label{Vpm}
V_\pm(\vec k,p)=g_s^2C_F\frac{-(m+k_\pm)+2m}{\sqrt{m^2+\kappa^2}
  (k_\pm^2-k^2)}=\frac{g_s^2C_F(\sqrt{m^2+\kappa^2}\mp 2m)}{2m
  \sqrt{m^2+\kappa^2}(\sqrt{m^2+\kappa^2}\mp m)}.
\end{equation}
For $m\ll\mu_f$ the restriction of the three-dimensional integral allows for
an expansion in $\kappa/m$. We obtain
\begin{eqnarray}
V_+(\vec k,p)&=&-4\pi\alpha_sC_F\left\{\frac1{\kappa^2}-\frac3{4m^2}
  +O\pfrac{\kappa^2}{m^4}\right\},\nonumber\\
V_-(\vec k,p)&=&-4\pi\alpha_sC_F\left\{\phantom{\frac1{\kappa^2}}-\frac3{4m^2}
  +O\pfrac{\kappa^2}{m^4}\right\}
\end{eqnarray}
and therefore
\begin{equation}
V(\vec k,p)\ =\ -4\pi\alpha_sC_F\left\{\frac1{\kappa^2}-\frac3{2m^2}
  +O\pfrac{\kappa^2}{m^4}\right\}.
\end{equation}
The first term is the Coulomb potential for a quark-antiquark interaction. The
second term can be related to the Breit-Fermi potential of the quark-antiquark
interaction~\cite{Landau} by summing over the spin states of the tensor
product of the final quark and the final antiquark state and using the same
kinematic constraints. Moreover, one can identify $V_+(\vec k,p)$ with the
scattering potential and $V_-(\vec k,p)$ with the annihilation potential.

One can integrate this radial symmetric potential over the space
components and end up with an expansion in $1/m$. Note that there are no
contributions to odd powers of $1/m$. But one can also integrate the exact
expressions for $V_\pm(\vec k,p)$ in Eq.~(\ref{Vpm}) up to the factorization
scale $\mu_f$ and obtain
\begin{equation}
\Sigma_{\rm soft}(\mu_f)=\frac{\alpha_sC_F}{2\pi} m
  \left\{3\ln\left(\frac{\mu_f}m+\sqrt{\frac{\mu_f^2}{m^2}+1}\,\right)
  -\frac{\mu_f}m\sqrt{\frac{\mu_f^2}{m^2}+1}\,\right\}.
\end{equation}
The expansion of this expression in small values of $\mu_f/m$ results in
\begin{equation}
\Sigma_{\rm soft}(\mu_f)=\frac{\alpha_sC_F}{\pi}\mu_f
  \left\{1-\frac{\mu_f^2}{2m^2}\right\}.
\end{equation}
The first term reproduces the result given in Ref.~\cite{Beneke} to
leading order in $\alpha_s$ while the second term is the recoil correction to
the static limit in this order of perturbation theory. This second term is
related to the Breit-Fermi potential but does not coincide with it. 

\section{Two loop contributions}
\begin{figure}
\begin{center}
\epsfig{figure=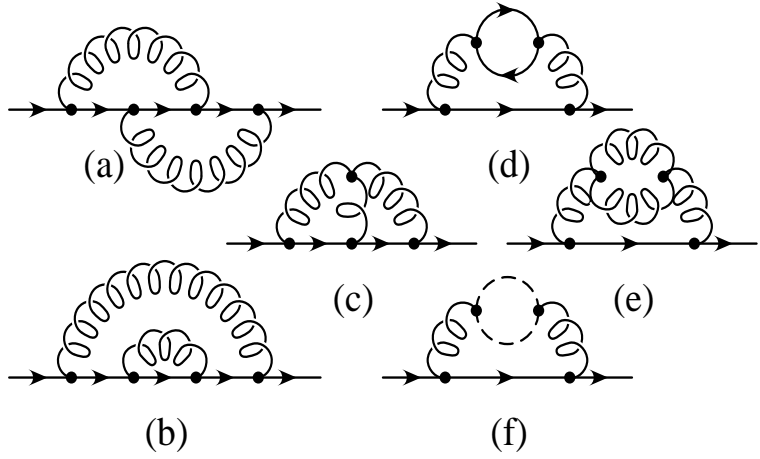, height=6truecm,width=10truecm}
\caption{\label{fig3}Two-loop contributions to the quark self energy}
\end{center}
\end{figure}

To take a step beyond the leading order perturbation theory, we consider
two-loop diagrams for the heavy quark self energy as shown in Fig.~\ref{fig3}.
We calculate them in Coulomb gauge, even though we stress that our final
result is gauge invariant. The gluon propagator in Coulomb gauge is given by
\begin{equation}
G^{ab}_{00}(k)=\frac{i\delta^{ab}}{\vec k\,^2},\qquad
G^{ab}_{ij}(k)=\frac{i\delta^{ab}}{k^2}
  \left(\delta_{ij}-\frac{k_ik_j}{\vec k\,^2}\right),\qquad i,j=1,2,3.
\end{equation}
The use of Coulomb gauge splits up the gluon propagators into a Coulomb term
(Coulomb gluon) and a transverse term (transverse gluon) where the first one
couples to the quark via the time components only. This splitting is shown in
Fig.~\ref{fig2}(b--d).

\subsection{The abelian diagrams}
\begin{figure}
\begin{center}
\epsfig{figure=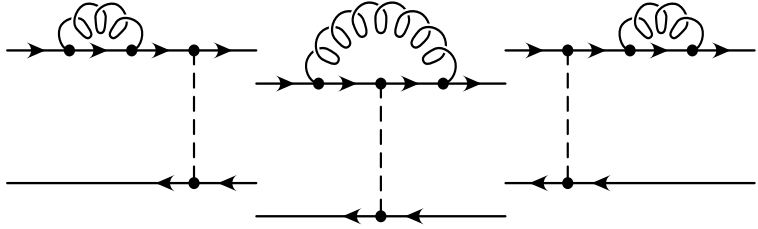, height=3truecm,width=10truecm}
\caption{\label{fig4}
Diagrams which cancel due to the classical Ward identity} 
\end{center}
\end{figure}

\begin{figure}
\begin{center}
\epsfig{figure=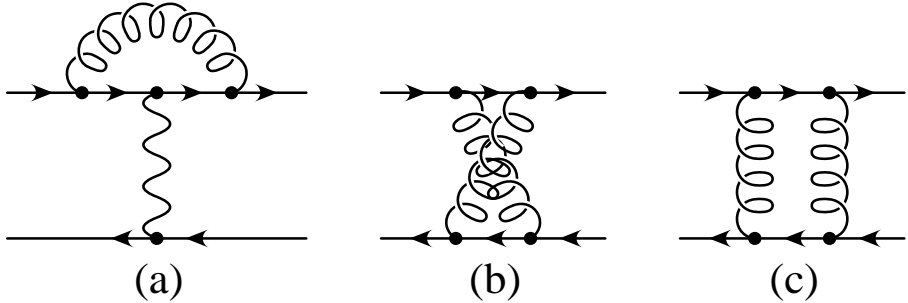, height=4truecm,width=12truecm}
\caption{\label{fig5}Diagrams which contribute to order $O(\mu_f^2/m^2)$}
\end{center}
\end{figure}

We start our consideration with the abelian diagrams shown in
Figs.~\ref{fig3}(a) and~(b). In cutting the quark line in all possible ways we
obtain a lot of diagrams. However, we find that the final contribution of
these diagrams to the soft part of the self energy is suppressed by
$\mu_f^2/m^2$. There can be found different arguments for this suppression.
First, in applying the classical Ward identity, the QED diagrams shown in
Fig.~\ref{fig4} cancel exactly at $|\vec k|\to 0$, the remaining contribution
is of order $O(\vec k^2/m^2)$. Concerning this note that the Ward identity for
the interaction vertex of a Coulomb gluon with the quark holds even in
non-abelian theories~\cite{Feinberg}. 
A second argument is that the interaction between a transverse gluon and a
non-relativistic quark as shown in Fig.~\ref{fig5}(a) is suppressed by
$\mu_f/m$, leading to an overall $\mu_f^2/m^2$ suppression. In addition, the
box diagrams in Figs.~\ref{fig5}(b) and~(c) are either suppressed by a factor
$\mu_f^2/m^2$ or give an iteration of the leading order potential. To
summarize, the diagrams in Figs.~\ref{fig3}(a) and~(b) give contributions only
of the order $g_s^4\mu_f^2/m^2$.

\subsection{The vacuum polarization of the gluon}
Following these arguments, it turns out that the only abelian diagrams which
can give a non-suppressed contribution to the soft part of the quark self
energy are the diagrams containing the vacuum polarization of the gluon as
shown in Fig.~\ref{fig3}(d--f). The simple calculation of these diagrams
within the $\overline{\rm MS}$ scheme, accounting only for light fermion
loops, gluon loop (and ghost loop if Feynman gauge is used) results after
renormalization in
\begin{eqnarray}
\Sigma_{\rm soft}^A&=&-\frac12\int^\mu\frac{d^3k}{(2\pi)^3}
  \left(-\frac{4\pi\alpha_s(\mu)C_F}{|\vec k|^2}\right)\
  \times\nonumber\\&&\times\ \left\{1+\frac{\alpha_s(\mu)}{4\pi}
  \left(\frac{31C_A}9-\frac{20T_FN_F}9-\left(\frac{11C_A}3-\frac{4T_FN_F}3
  \right)\ln\pfrac{|\vec k|^2}{\mu^2}\right)\right\}\nonumber\\
  &=&\frac{\alpha_S(\mu)C_F}\pi\mu_f\left\{1+\frac{\alpha_s(\mu)}{4\pi}
  \left(a_1-\beta_0\ln\pfrac{\mu_f^2}{\mu^2}\right)\right\}.
\end{eqnarray}
This result has been anticipated because the expression in the curly
brackets of the integrand reproduces 
the next-to-leading order correction to the QCD Coulomb potential. 
 
\subsection{The non-abelian diagram}
\begin{figure}
\begin{center}
\epsfig{figure=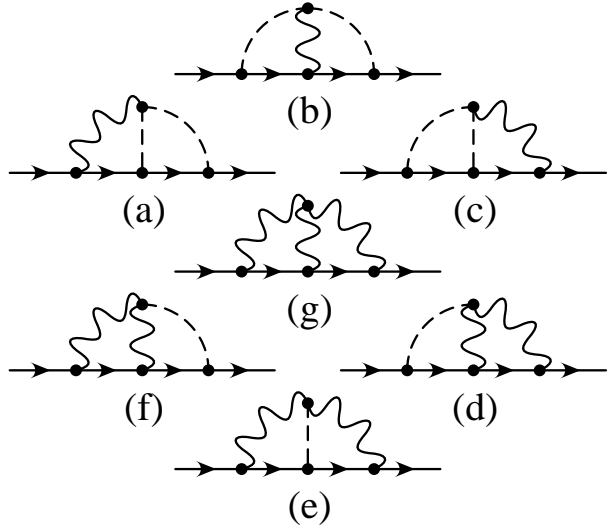, height=7truecm,width=8truecm}
\caption{\label{fig6}The non-abelian self energy diagram in Coulomb gauge;
displayed are Coulomb gluons (wiggles) and transverse gluons (dashed lines)}
\end{center}
\end{figure}

\begin{figure}
\begin{center}
\epsfig{figure=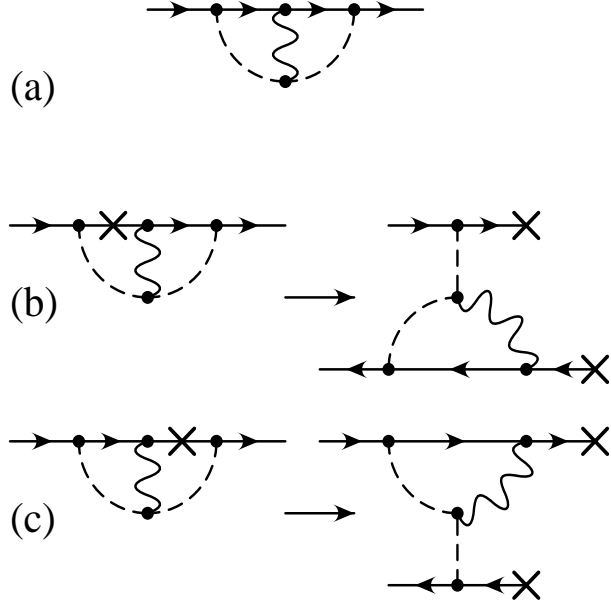, height=8truecm,width=8truecm}
\caption{\label{fig7}The soft part of the non-abelian diagram under
  consideration}
\end{center}
\end{figure}

In this subsection we calculate the non-abelian diagram shown in
Fig.~\ref{fig3}(c). In Coulomb gauge this diagram gives rise to seven two-loop
diagrams which are shown in Fig.~\ref{fig6}.
Direct calculations show that only the diagram in Fig.~\ref{fig6}(b) gives
a contribution of order $g_s^4\mu_f/m$ while the other diagrams are of order
$g_s^4\mu_f^2/m^2$ or vanish to this order in the coupling after applying the
renormalization procedure (see e.g.\ Ref.~\cite{KummerMoedritsch}). 
The calculation of the diagram in Fig.~\ref{fig6}(b) is simple 
and we show it in detail. The
contribution of this diagram to the self energy is given by
\begin{eqnarray}
-i\Sigma^{6b}(\slp)&=&\int\frac{d^4k_1}{(2\pi)^4}
  \frac{d^4k_2}{(2\pi)^4}\frac{i}{\vec k_1^2}\frac{i}{\vec k_2^2}
  \frac{i}{(k_1-k_2)^2}\left(\delta_{ij}
  -\frac{(k_1-k_2)_i(k_1-k_2)_j}{(\vec k_1-\vec k_2)^2}\right)
  g_sf_{abc}(k_1+k_2)^jg^{00}\times\nonumber\\&&\times\
  \frac{1+\gamma^0}2(-ig_s\gamma^0T_a)\frac{i}{\slp+\slk_2-m}(-ig_s\gamma^iT_b)
  \frac{i}{\slp+\slk_1-m}(-ig_s\gamma^0T_c)\frac{1+\gamma^0}2.
\end{eqnarray}
Here the Lorentz structure of the three-gluon vertex is reduced to $k_1+k_2$.
The Dirac structure of the integrand can be simplified to
\begin{equation}
\frac{1+\gamma^0}2\left\{-(2m+k_{20})\gamma^i\vec k_1\vec\gamma
  -\vec k_2\vec\gamma\gamma^i(2m+k_{10})\right\}.
\end{equation}
The rest of the diagram is symmetric with respect to the interchange of the
two line momenta $k_1$ and $k_2$. That provides us with 
a further simplification of
the Dirac structure, the self energy reduces to
\begin{equation}
\Sigma^{6b}(\slp)=g_s^4C_FC_A\int\frac{d^4k_1}{(2\pi)^4}\frac{d^4k_2}{(2\pi)^4}
  \frac{(4m+k_{10}+k_{20})\left(\vec k_1^2\vec k_2^2-(\vec k_1\vec k_2)^2
  \right)}{\vec k_1^2\vec k_2^2(\vec k_1-\vec k_2)^2(k_1-k_2)^2((p+k_1)^2-m^2)
  ((p+k_2)^2-m^2)}
\end{equation}
where $T_aT_bT_cf_{abc}=iC_FC_A/2$ has been used. We now employ 
the replacements
\begin{equation}
\frac1{(p+k_1)^2-m^2}\rightarrow-i\pi\delta((p+k_1)^2-m^2),\qquad
\frac1{(p+k_2)^2-m^2}\rightarrow-i\pi\delta((p+k_2)^2-m^2)
\end{equation}
i.e.\ we set cuts at the two intermediate quark lines separately to obtain
the two parts $\Sigma^{6b}_{{\rm soft}1}$ and $\Sigma^{6b}_{{\rm soft}2}$ 
of the soft contribution $\Sigma^{6b}_{\rm soft}$ of the self energy, as shown
in Fig.~\ref{fig7}.
Taking the first cut, the delta function removes the integration over
$k_{10}$. At the same time our definition of the soft contribution imposes a
restriction $|\vec k_1|<\mu_f$ on the space components of the first line
momentum. The delta function
\begin{equation}
\delta\left((p+k_1)^2-m^2\right)=\frac1{2\sqrt{m^2+\vec k_1^2}}
  \left(\delta(k_{10}-k_{1+})+\delta(k_{10}-k_{1-})\right)
\end{equation}
with
\begin{equation}
k_{1\pm}=\pm\sqrt{m^2+\vec k_1^2}-m
\end{equation}
causes two contributions which are known as scattering and annihilation
amplitude (according to $k_{1+}$ and $k_{1-}$, resp.). The integration over
$k_{20}$ is done by using the residue theorem. Actually there are only two
denominator factors which can contribute to poles of the integrand, namely
$(k_1-k_2)^2$ and $((p+k_2)^2-m^2)$. In summing up the four contributions
from the integration over $k_{20}$ and $k_{10}$ we end up with a two-fold
three-dimensional integral over the space components of the two line momenta
where the first integral is restricted by $|\vec k_1|<\mu_f$ as mentioned
earlier. We now impose the restriction $|\vec k_1|<\mu_f$ on the integrand to
simplify it and obtain
\begin{equation}
\Sigma^{6b}_{{\rm soft}1}=\frac{g_s^4C_FC_A}{4m}\int^{\mu_f}
  \frac{d^3k_1}{(2\pi)^3}\int\frac{d^3k_2}{(2\pi)^3}\frac{\vec k_1^2\vec k_2^2
  -(\vec k_1\vec k_2)^2}{\vec k_1^2(\vec k_2^2)^2(\vec k_1-\vec k_2)^2}.
\end{equation}
The integral over the space components of $k_2$ can be easily done by
executing the angular integration followed by the radial integration. We obtain
\begin{equation}
\int\frac{d^3k_2}{(2\pi)^3}\frac{\vec k_1^2\vec k_2^2
  -(\vec k_1\vec k_2)^2}{\vec k_1^2(\vec k_2^2)^2(\vec k_1-\vec k_2)^2}
  =\frac1{16|\vec k_1|}
\end{equation}
and therefore, finally
\begin{equation}
\Sigma^{6b}_{{\rm soft}1}=\frac{\alpha_s^2C_FC_A}{16m}\mu_f^2.
\end{equation}
Symmetry considerations show that $\Sigma^{6b}_{{\rm soft}2}$ gives exactly
the same contribution. As mentioned before, there are no other non-abelian
contributions, therefore we obtain
\begin{equation}
\Sigma_{\rm soft}^{NA}=\frac{\alpha_s^2C_FC_A}{8m}\mu_f^2.
\end{equation}
This result has been anticipated, too, to be minus one half of the non-abelian
correction to the QCD Coulomb potential, which is known in the literature 
(see for example Refs.~\cite{Gupta,Duncan}),
\begin{eqnarray}
\Sigma_{\rm soft}^{NA}&=&-\frac12\int^{\mu_f}\frac{d^3k}{(2\pi)^3}
  \left\{-\frac{\pi^2\alpha_s^2C_FC_A}{m|\vec k|}\right\}
  \ =\ \frac{\alpha_s^2C_FC_A}{8m}\mu_f^2.
\end{eqnarray}
This calculation concludes the considerations of the two-loop diagrams shown in
Fig.~\ref{fig3}.

\subsection{Our final result}
Summarizing all contribution up to NNLO accuracy, we obtain
\begin{equation}\label{psmass}
m_{\overline{\rm PS}}(\mu_f)-m=-\frac{\alpha_s(\mu)C_F}\pi\mu_f
  \left\{1+C'_0\frac{\mu_f}m+C''_0\frac{\mu_f^2}{m^2}
  +\frac{\alpha_s(\mu)}{4\pi}\left(C_1+C'_1\frac{\mu_f}m\right)
  +C_2\pfrac{\alpha_s(\mu)}{4\pi}^2\right\} 
\end{equation}
where $m$ is the pole mass, $\mu$ is the renormalization scale. This result is
considered to be of the order $O(\alpha_s^2)$ because we imply that the ratio
$\mu_f/m$ is typically of the order of $\alpha_s$ or smaller. The scale
$\mu_f$ is the factorization scale, and
\begin{eqnarray}
C_0&=&1,\qquad C'_0\ =\ 0,\qquad C''_0\ =\ -\frac12,\nonumber\\
C_1&=&a_1-2\beta_0\ln\pfrac{\mu_f}\mu,\qquad
  C'_1\ =\ C_A\frac{\pi^2}2,\nonumber\\
C_2&=&a_2-2(2a_1\beta_0+\beta_1)\left(\ln\pfrac{\mu_f}\mu-1\right) 
  +4\beta_0^2\left(\ln^2\pfrac{\mu_f}\mu-2\ln\pfrac{\mu_f}\mu+2\right).
\end{eqnarray}
The constants $a_1$, $a_2$, $\beta_0$, and $\beta_1$ are given in Appendix A.
The coefficients $C_1$ and $C_2$ have been derived in Ref.~\cite{Beneke} by
using known corrections to the QCD potential. In this work we have derived the
coefficients $C'_0$, $C''_0$, and $C'_1$. Note that our result can be
represented in a condensed form as 
\begin{equation}
m_{\overline{\rm PS}}(\mu_f)-m=-\frac12\int^{\mu_f}\frac{d^3k}{(2\pi)^3}
  \left(V_C(|\vec k|)+ V_R(|\vec k|)+V_{NA}(|\vec k|)\right)
\end{equation}
where the first term $V_C$ is the static Coulomb potential, $V_R$ is the
relativistic correction (which is related to Breit-Fermi potential but does
not coincide with it), and $V_{NA}$ is the non-abelian correction.

\section{Top quark production near threshold\\ with NNLO accuracy}
In this section we consider the cross section of the process
$e^+e^-\to t\bar t$ in the near threshold region where the velocity $v$ of
the top quark is small. It is well-known that the conventional perturbative 
expansion does not work in the non-relativistic region because of the presence
of the Coulomb singularities at small velocities $v\to 0$. The terms
proportional to $(\alpha_s/v)^n$ appear due to the instantaneous Coulomb
interaction between the top and the antitop quark. The standard technique for
resumming the terms $(\alpha_s/v)^n$ consists in using the Schr\"odinger
equation for the Coulomb potential and to find the Green 
function~\cite{Fadin:1987wz,Fadin:1988fn}. The Green function is then related
to the cross section by the optical theorem. In order to determine NLO
corrections to the cross section we need to know the short-distance correction
to the vector current~\cite{NLOcurrent}, the NLO correction to the Coulomb
potential~\cite{Billoire:1980ih}, and the contribution of the non-factorizable
corrections~\cite{Melnikov:1994np,Fadin:1994dz,Fadin:1994kt}. It was proven
that the non-factorizable corrections cancel in the inclusive cross section at
NLO and beyond~\cite{Melnikov:1994np,Fadin:1994dz,Fadin:1994kt}.  

At NNLO the situation is more complicated. One of the obstacles for a
straightforward calculation are the UV divergences coming from relativistic
corrections to the Coulomb potential. This problem can be solved by a proper
factorization of the amplitudes and by employing effective theories. We want
to sketch the derivation of the inclusive cross section. The inclusive cross
section can be obtained from the correlation function of two vector currents
$j_{\mu}(x)=\bar t(x)\gamma_\mu t(x)$,
\begin{equation}
\Pi_{\mu\nu}(p^2)=i\int d^4xe^{ip\cdot x}
  \langle 0|T\{j_\mu(x),j_\nu(0)\}|0\rangle. 
\end{equation}
The first step is to expand the Lagrangian and the currents
\begin{equation}
j_i=\bar t\gamma_it=c_1\psi^\dagger\sigma_i\chi
  -\frac{c_2}{6m^2}\psi^\dagger\sigma_i(i{\vec D})^2\chi , 
\end{equation}
consistently in $1/m$. The useful language for treating the one and two
non-relativistic quark(s) is provided by the
NRQCD~\cite{Caswell:1986ui,Bodwin:1995jh} and the PNRQCD~\cite{Pineda2},
respectively. After the expansion, the relative cross section reads
\begin{equation}\label{cross}
R=\sigma(e^+e^-\to t\bar t)/\sigma_{pt}
  =e^2_QN_c\frac{24\pi}sC(r_0){\rm Im}\left[\left(1-\frac{\vec p\,^2}{3m^2}
  \right)G(r_0,r_0|E+i\Gamma)\right]\Bigg|_{r_0\to 0}
\end{equation}
where $\sigma_{pt}=4\pi\alpha^2/3s$, $e_Q$ is the electric charge of the top
quark, $N_c$ is the number of colors, $\sqrt{s}=2m+E$ is the total
center-of-mass energy of the quark-antiquark system, $m$ is the top quark pole
mass and $\Gamma$ is the top quark width. The unknown coefficient $C(r_0)$ can
be fixed by using a direct QCD calculation of the vector vertex at NNLO in the
so-called intermediate region~\cite{Czarnecki:1998vz,Beneke:1998zp} and by
using the direct matching procedure suggested in Ref.~\cite{HoangM}. The result
of such a matching procedure for the coefficient $C(r_0)$ is given in the
Appendix~\cite{Hoang:1998xf,Melnikov:1998pr,Yakovlev:1999ke,Beneke:1999qg}.

The function  $G(\vec r, \vec r\,'|E+i\Gamma)$ is the non-relativistic 
Green function. It satisfies the Schr\"odin\-ger equation
\begin{equation}
(H-E-i\Gamma)G(\vec r,\vec r\,'|E+i\Gamma)=\delta(\vec r-\vec r\,').
\end{equation}
The non-relativistic Hamiltonian of the heavy quark-antiquark system reads
\begin{equation}\label{www}
H=H_0+W(r),\qquad H_0=\frac{\vec p\,^2}m+V_C(r), 
\end{equation}
where $V_C(r)$ is the static QCD potential of the heavy quark-antiquark system
at NNLO order,
\begin{eqnarray}
V_C(r)&=&-\frac{\alpha_s(\mu)C_F}{r}\Bigg\{1+\frac{\alpha_s(\mu)C_F}{4\pi}
  \left(2\beta_0\ln(\mu'r)+a_1\right)\nonumber\\&&
  +\pfrac{\alpha_s(\mu)}{4\pi}^2\left(\beta_0^2
  \left(4\ln^2(\mu'r)+\frac{\pi^2}3\right)+2(\beta_1+2\beta_0a_1)\ln(\mu'r)
  +a_2\right)\Bigg\},
\end{eqnarray}
$\mu'=\mu e^{\gamma_E}$, $\mu$ is the renormalization scale, and $\gamma_E$ is
Euler's constant. The color factors are given by $C_F=4/3$, $C_A=3$ and
$T_F=1/2$. The number of light quark flavors is $N_F = 5$. The coefficients
$a_1$ and $a_2$ were calculated in~\cite{Fischler}
and~\cite{Peter,Schroder}, respectively, and are listed in the Appendix.

The function $W(r)$ in Eq.~(\ref{www}) is the QCD generalization of the QED
Breit-Fermi Hamiltonian~\cite{Landau,Gupta,Duncan}. We consider here the
quark-antiquark production in the $S$-wave mode. The Breit-Fermi Hamiltonian
for the final state with $\vec L=0$, $\vec S^2=2$ reads 
\begin{equation}\label{HBF1}
W(r)=-\frac{\vec p\,^4}{4m^3}+\frac{11\pi C_F\alpha_s(\mu)}{3m^2}\delta(\vec r)
  -\frac{C_F\alpha_s(\mu)}{2m^2}\left\{\frac{1}{r},\vec p\,^2\right\}
  -\frac{C_AC_F\alpha_s^2(\mu)}{2mr^2}.
\end{equation}
It has been demonstrated in 
Refs.~\cite{Hoang:1998xf,Melnikov:1998pr,Yakovlev:1999ke} that the
Schr\"odinger equation can be reduced to the equation
\begin{equation}\label{se1}
(H_1-E_1)G_1(r,r'|E_1)=\delta^3(r-r')
\end{equation}
with the energy $E_1=\bar E+\bar E^2/4m$, $\bar E=E+i\Gamma$, and with the
Hamiltonian 
\begin{equation}
H_1=\frac{\vec p\,^2}{m}+V_C(r)+\frac{3\bar E}{2m}V_0(r)
  -\left(\frac23+\frac{C_A}{C_F}\right)\frac{V_0^2(r)}{2m},\qquad
V_0(r)=-\frac{\alpha_s(\mu)C_F}r.
\end{equation}
The final expression for the NNLO cross section is given by
\begin{equation}\label{final}
R^{\rm NNLO}(E)=\frac{8\pi}{m^2}\left\{1-4C_F\frac{\alpha_s(m)}\pi
  +C_2(r_0)C_F\pfrac{\alpha_s(m)}\pi^2\right\}
  {\rm Im}\left[\left(1-\frac{5\bar E}{6m}\right)G_1(r_0,r_0|E_1)\right]
\end{equation}
with $G_1(r_0,r_0|E_1)$ being the solution of Eq.~(\ref{se1}) at $r=r'=r_0$
while $C_2(r_0)$ is taken from Eqs.~(\ref{sdc}--\ref{sdc2}). For the numerical
solution we use the program derived in Ref.~\cite{Yakovlev:1999ke} by
one of the authors.

\subsection{Relations between the masses}
Our main result is Eq.~(\ref{psmass}), it enables us to relate the pole mass to
the $\overline{\rm PS}$ mass or the PS mass (the latter by leaving out the
terms of higher order in $1/m$). Note that we did not include electroweak
corrections neither in this mass relation nor in the cross section. To relate
the pole mass to the $\overline{\rm MS}$ mass we use the three-loop relation
which has been derived in Ref.~\cite{MelRit,ChetyrkinSteinhauser},
\begin{eqnarray}\label{msmass}
\frac{m_{\rm pole}}{\overline{m}(\overline{m})}&=&1+\frac43\pfrac{\alpha_s}\pi
  +\pfrac{\alpha_s}\pi^2(-1.0414N_F+13.4434)\nonumber\\&&
  +\pfrac{\alpha_s}\pi^3(0.6527N_F^2-26.655N_F+190.595)
\end{eqnarray}
where $\alpha_s=\alpha_s(\overline{m})$ is taken at the $\overline{\rm MS}$
mass. More precisely, we fix the $\overline{\rm MS}$ mass to take the values
$\overline{m}(\overline{m})=160\GeV$, $165\GeV$, and $170\GeV$ and use
Eq.~(\ref{msmass}) to determine the pole mass at LO, NLO, and NNLO. This pole
mass is then used as input parameter $m$ in Eq.~(\ref{psmass}) to determine
the PS and $\overline{\rm PS}$ masses at LO, NLO, and NNLO. The obtained values
for the PS and the $\overline{\rm PS}$ mass differ only in NNLO. The results
of these calculations are collected in Table~\ref{tab1}, together with the
estimates for ``large $\beta_0$'' accuracy~\cite{AgliettiLigeti,BenekeBraun}.
Note that the same values for the $\overline{\rm MS}$ mass have been used for
Tables~2 and~3 in Ref.~\cite{Review}. The obtained mass values can now be used
for an analysis of the Schr\"odinger equation.

Taking the $\overline{\rm PS}$ mass instead of the PS mass, we observe an
improvement of the convergence. The differences for the mass values in going
from LO to NLO to NNLO to the ``large $\beta_0$'' estimate for e.g.\
$\overline{m}(\overline{m})=165\GeV$ read $7.64\GeV$, $1.64\GeV$, $0.52\GeV$,
and $0.25\GeV$ for the pole mass, $6.72\GeV$, $1.21\GeV$, $0.29\GeV$, and
$0.08\GeV$ for the PS mass and finally $6.72\GeV$, $1.21\GeV$, $0.27\GeV$, and
$0.08\GeV$ for the $\overline{\rm PS}$ mass.

\begin{table}[t]
\begin{center}
\begin{tabular}{|c||c|c|c|c||c|c|c|c|}\hline&&&&&&&&\\
$\overline{m}(\overline{m})$&$m_{\rm PS}^{\rm LO}$&$m_{\rm PS}^{\rm NLO}$&
$\displaystyle\matrix{m_{\rm PS}^{\rm NNLO}\cr
  m_{\overline{\rm PS}}^{\rm NNLO}}$&
$\displaystyle\matrix{m_{\rm PS}^{\beta_0}\cr
  m_{\overline{\rm PS}}^{\beta_0}}$&$m_{\rm pole}^{\rm LO}$&
  $m_{\rm pole}^{\rm NLO}$&
  $m_{\rm pole}^{\rm NNLO}$&$m_{\rm pole}^{\beta_0}$\\&&&&&&&&\\\hline
\hline &&&&&&&&\\
  160.0&166.51&167.69&$\displaystyle\matrix{167.97\cr 167.95}$&
  $\displaystyle\matrix{168.05\cr 168.03}$&167.44&169.05&169.56&169.80\\
\hline &&&&&&&&\\
  165.0&171.72&172.93&$\displaystyle\matrix{173.22\cr 173.20}$&
  $\displaystyle\matrix{173.30\cr 173.28}$&172.64&174.28&174.80&175.05\\
\hline &&&&&&&&\\
  170.0&176.92&178.17&$\displaystyle\matrix{178.47\cr 178.45}$&
  $\displaystyle\matrix{178.55\cr 178.53}$&177.84&179.52&180.05&180.30\\
\hline
\end{tabular}
\caption{\label{tab1}
Top quark mass relations for the $\overline{\rm MS}$, PS, $\overline{\rm PS}$,
and the pole mass at LO, NLO, NNLO, and ``large $\beta_0$'' accuracy in $\GeV$.
We fix the $\overline{\rm MS}$ mass to be $\overline{m}(\overline{m})=160\GeV$,
$165\GeV$, and $170\GeV$ and find the pole mass at LO, NLO, and NNLO from the
three-loop relation in Eq.~(\ref{msmass}). The PS and $\overline{\rm PS}$
masses are derived from the pole mass by using Eq.~(\ref{psmass}) (without or
with the $1/m$ contributions, respectively). We use the QCD coupling constant
$\alpha_s(m_Z)=0.119$ \cite{PDG},  $\mu=\overline{m}(\overline{m})$, and
$\mu_f=20\GeV$ for the factorization scale in the PS and $\overline{\rm PS}$
masses.}
\end{center}
\end{table}

\begin{figure}
\centerline{\epsfig{file=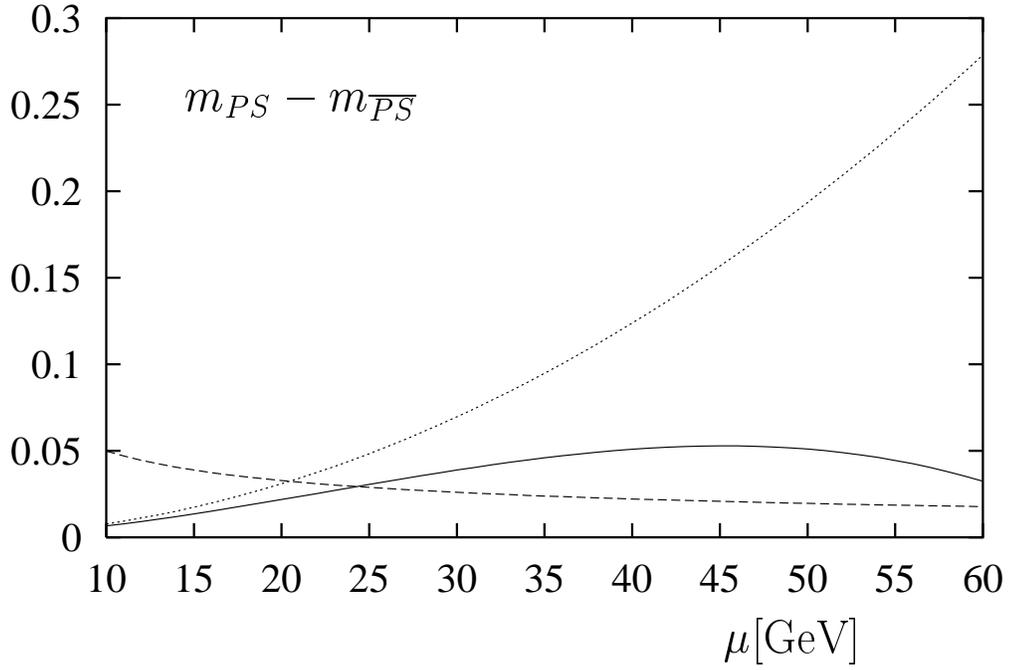,scale=0.85}}
\caption{\label{fig8}The difference between the PS and the $\overline{\rm PS}$
mass (in GeV) as a function of the factorization scale $\mu_f$ (solid line) at
$\mu=15\GeV$. The dotted line shows only the non-abelian part of the
difference. The dependence of $m_{\overline{\rm PS}}-m_{\rm PS}$ as a function
of the normalization scale $\mu$ at $\mu_f=30\GeV$ is shown as dashed line.}
\end{figure}

In Fig.~\ref{fig8} we show the difference between the $\overline{\rm PS}$ and
the PS mass (in GeV) as a function of the factorization  scale $\mu_f$ (solid
line) at $\mu=15\GeV$. It is interesting to observe that the non-abelian part
of the difference between the $\overline{\rm PS}$ and the PS mass
(dotted line in Fig.~\ref{fig8}) can be as large as $200\MeV$. But the recoil
correction cancel in part the non-abelian one and therefore the final
difference is not more than $50\MeV$. The dependence of
$m_{\overline{\rm PS}}-m_{\rm PS}$ on the renormalization scale at
$\mu_f=30\GeV$ is given in Fig.~\ref{fig8} by the dashed curve.

\subsection{The concept with the pole mass}
The top quark cross section at LO, NLO, and NNLO is shown in Fig.~\ref{fig9}
as a function of the center-of-mass energy. For the top quark pole mass we use
$m_t=175.05\GeV$, for the top quark width $\Gamma_t=1.43\GeV$, and for the QCD
coupling constant $\alpha_s(m_Z)=0.119$~\cite{PDG}. Different values
$\mu=15\GeV$, $30\GeV$, and $60\GeV$ for the renormalization scale are
selected because they roughly correspond to the typical spatial momenta for
the top quark. For solving the Schr\"odinger equation we use the program
written by one of the authors~\cite{Yakovlev:1999ke}. Note that we do not take
into account an initial photon radiation which would change the shape of the
cross section. This can be easily included in the Monte Carlo simulation.

The NNLO curve modifies the line shape by the amount of $20-30\%$ which is as
large as the NLO correction. It also shifts the position of the $1S$ peak by
approximately $600\MeV$. These large shifts of the peak position were expected.
As we discussed above (and is well-known in the
literature~\cite{Beneke:1994sw,Vainshtein}), the pole mass definition suffers
from the renormalon ambiguity. The top quark pole mass cannot be defined better
than $O(\Lambda_{\rm QCD})$. Large NNLO corrections and a large shift of the
$1S$ resonance can spoil the top quark mass measurement at the NLC.

\subsection{The concept with the PS mass}
In this subsection we discuss the calculation concept using the PS mass. We
redefine the pole mass through the PS mass using the relation given in
Eq.~(\ref{psmass}) without the $1/m$ contributions and then use the PS mass as
an input parameter for our numerical analysis at LO, NLO, and NNLO. In
Fig.~\ref{fig10} we show the top quark cross section expressed in terms of
$m_{\rm PS}(\mu_f)$ at LO, NLO, and NNLO (like in Fig.~\ref{fig9}) as a
function of the center-of-mass energy. We take
$m_{\rm PS}(\mu_f=20\GeV)=173.30\GeV$ which corresponds to the pole mass
$m=175.05\GeV$ according to Table~\ref{tab1}. In looking at Fig.~\ref{fig10}
we observe an improvement in the stability of the position of the first peak
in comparison to the previous analysis as we go from LO to NLO to NNLO.
Actually, for the three values $\mu=15\GeV$, $30\GeV$, and $60\GeV$ we obtain
the maxima of the $1S$ peak for NNLO at $s_{\rm max}=347.32\GeV$, $347.41\GeV$,
and $347.48\GeV$ while the maximal values are given by $R_{\rm max}=1.379$,
$1.184$, and $1.088$, respectively. This demonstrated that a large variation
in the renormalization scale $\mu$ gives rise only to a shift of about
$160\MeV$ for the $1S$ peak position at NNLO while the variation for
$R_{\rm max}$ is still large.

\subsection{The concept with the $\overline{\rm PS}$ mass}
\begin{figure}
\centerline{\epsfig{file=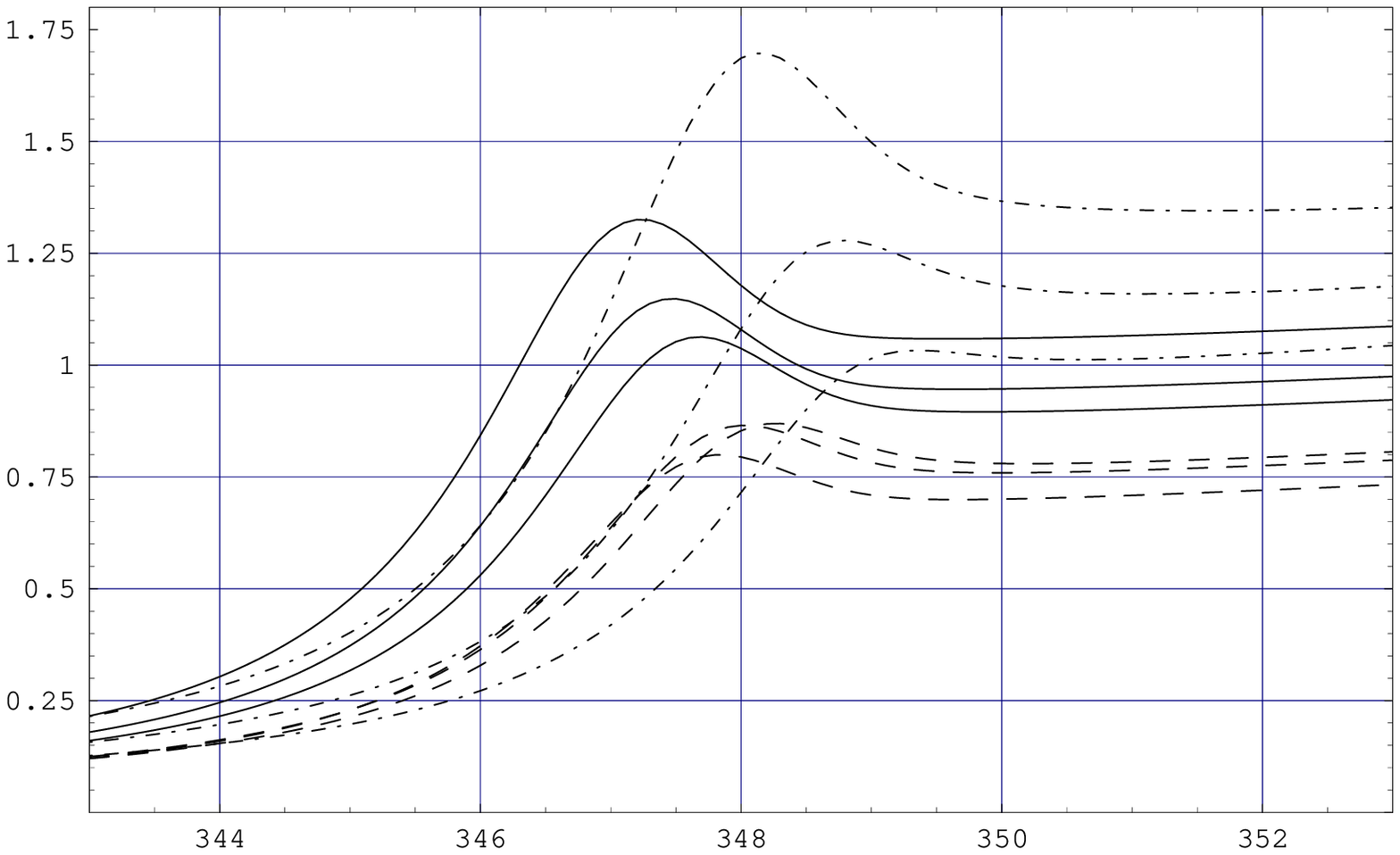,scale=0.85}}
\caption{\label{fig9}The concept using the pole mass:
shown is the relative cross section $R(e^+e^-\to t\bar t)$ as a function of
the center-of-mass energy in $\GeV$ for the LO (dashed-dotted lines), NLO
(dashed lines), and NNLO (solid lines) approximation. We take the value
$m_t=175.05\GeV$ for the pole mass of the
top quark, $\Gamma_t=1.43\GeV$ for the top quark width,
$\alpha_s(m_Z)=0.119$ and different values
$\mu=15\GeV$, $30\GeV$, and $60\GeV$ for the renormalization scale.}
\end{figure}

\begin{figure}
\centerline{\epsfig{file=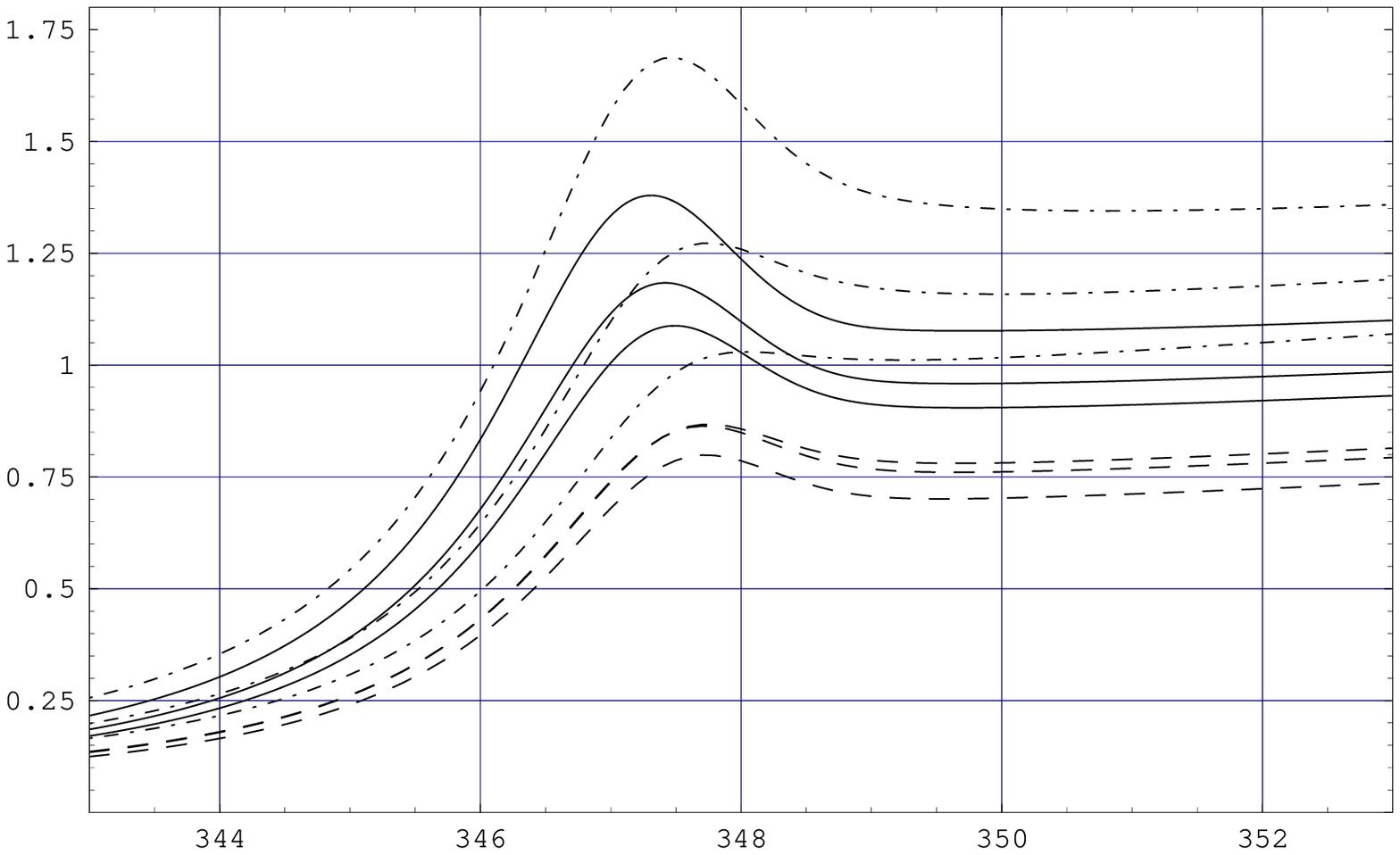,scale=0.85}}
\caption{\label{fig10}The concept using the PS mass:
shown is the relative cross section $R(e^+e^-\to t\bar t)$ as a function of
the center-of-mass energy in $\GeV$ for the LO (dashed-dotted lines), NLO
(dashed lines), and NNLO (solid lines) approximation. We take the value
$m_{\rm PS}=173.30\GeV$ for the PS mass of the
top quark, $\Gamma_t=1.43\GeV$ for the top quark width,
$\alpha_s(m_Z)=0.119$ and different values
$\mu=15\GeV$, $30\GeV$, and $60\GeV$ for the renormalization scale.}
\end{figure}

\begin{figure}
\centerline{\epsfig{file=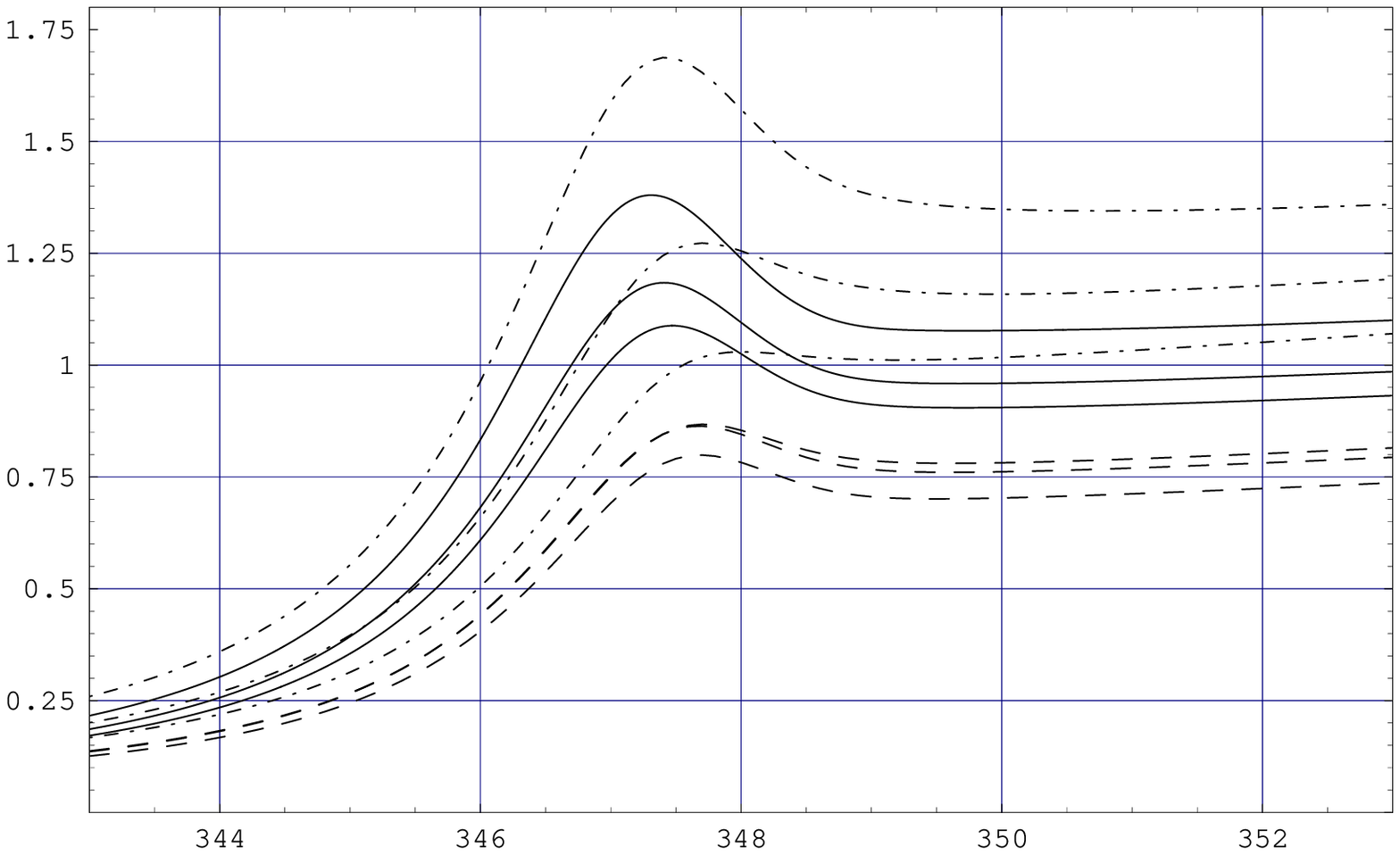,scale=0.85}}
\caption{\label{fig11}The concept using the $\overline{\rm PS}$ mass:
shown is the relative cross section $R(e^+e^-\to t\bar t)$ as a function of
the center-of-mass energy in $\GeV$ for the LO (dashed-dotted lines), NLO
(dashed lines), and NNLO (solid lines) approximation. We take the value
$m_{\overline{\rm PS}}=173.28\GeV$ for the $\overline{\rm PS}$ mass of the
top quark, $\Gamma_t=1.43\GeV$ for the top quark width,
$\alpha_s(m_Z)=0.119$ and different values
$\mu=15\GeV$, $30\GeV$, and $60\GeV$ for the renormalization scale.}
\end{figure}

In this subsection we discuss the calculation concept where we use the
$\overline{\rm PS}$ mass. We redefine the pole mass through the
$\overline{\rm PS}$ mass by using the relation given in Eq.~(\ref{psmass}) and
then use the $\overline{\rm PS}$ mass as an input parameter for the numerical
analysis at LO, NLO, and NNLO. In Fig.~\ref{fig11} we show the top quark cross
section expressed in terms of $m_{\overline{\rm PS}}(\mu_f)$ at LO, NLO, and
NNLO (like in Fig.~\ref{fig9}) as a function of the center-of-mass energy. We
take $m_{\overline{\rm PS}}(\mu_f=20\GeV)=173.28\GeV$. Again we observe a very
good stability of the position of the first peak as we go from LO to NLO to
NNLO, similar to the one observed for the PS mass case. Studying the size of
the NNLO corrections to the peak positions we conclude that the current
theoretical uncertainty of the determination of the PS mass from the $1S$ peak
position is about $100-200\MeV$.

\section{Conclusion and discussions}
We introduced the so-called $\overline{\rm PS}$ mass which is an
infrared-insensitive quark mass, naturally defined by subtracting the soft
part of the quark self energy from the pole mass. We demonstrated a deep
relation of this definition with the static quark-antiquark potential. At
leading order in $1/m$ the $\overline{\rm PS}$ mass coincides with the PS
mass, defined in a completely different manner. In contrast to the pole mass,
the $\overline{\rm PS}$ mass is not sensitive to the non-perturbative QCD
effects. Going beyond the static limit, the small normalization point
introduces recoil corrections to the relation of the pole mass with the
$\overline{\rm PS}$ mass which have been calculated in this paper. In
addition we demonstrated that, if we use the PS or the $\overline{\rm PS}$
mass in the calculations, the perturbative predictions for the cross section
become much more stable at higher orders of QCD (shifts are below $0.1\GeV$).
This understanding removes one of the obstacles for an accurate top mass
measurement and one can expect that the top quark mass will be extracted from
a threshold scan at NLC with an accuracy of about $100-200\MeV$.\\[12pt]
{\bf Acknowledgements:}
We are grateful to Ratindranath Akhoury, Ed Yao and Martin Beneke for valuable
discussions. O.Y.\ acknowledges support from the US Department of Energy (DOE).
S.G.\ acknowledges a grant given by the DFG, FRG, he also would like to thank
the members of the theory group at the Newman Lab for their hospitality during
his stay.

\section*{Appendix}
\setcounter{equation}{0}
\def\theequation{A\arabic{equation}}
The coefficients $a_1$ and $a_2$ were calculated in Refs.~\cite{Fischler}
and~\cite{Peter,Schroder}, respectively, and are given by
\begin{eqnarray}
a_1&=&\frac {31}9C_A-\frac{20}9T_FN_F,\nonumber\\
a_2&=&\left(\frac{4343}{162}+4\pi^2-\frac{\pi^4}4
  +\frac{22}3\zeta_3\right)C_A^2
  -\left(\frac{1798}{81}+\frac{56}3\zeta_3\right)C_AT_FN_F\nonumber\\&&
  +\left(\frac{20}9T_FN_F\right)^2
  -\left(\frac{55}3-16\zeta_3\right)C_FT_FN_F.
\end{eqnarray}
The first two coefficients in the expansion of the QCD $\beta$-function are
\begin{eqnarray}
\beta_0=\frac{11}3C_A-\frac43T_FN_F,\quad
\beta_1=\frac{34}3C_A^2-\frac{20}3C_AT_FN_F-4C_FT_FN_F.
\end{eqnarray}
The short distance coefficient reads
\begin{eqnarray}\label{sdc}
C(r)&=&1-4C_F\frac{\alpha_s(\mu_h)}\pi+
  C_2(r)C_F\pfrac{\alpha_s(\mu_h)}\pi^2,\nonumber\\
C_2(r)&=&A_1\log(r/a)+A_2\log(m/\mu_{h})+A_3
\end{eqnarray}
where the hard renormalization scale is taken to be the pole mass,
$\mu_h=m$, $a=e^{2-\gamma_E}/2m$, and
\begin{eqnarray}
A_1&=&\pi^2\left(C_A+\frac{2C_F}3\right),\qquad A_2\ =\ 2\beta_0,\nonumber\\
A_3&=&C_2^AC_F+C_2^{NA}C_A+C_2^LT_FN_F+C_2^HT_F
\end{eqnarray}
where $N_F=5$ is the number of the light flavors, and
\begin{eqnarray}\label{sdc2}
C_2^A&=&\frac{39}4-\zeta_3+\pi^2\left(\frac43\ln2-\frac{35}{18}\right),
  \nonumber\\
C_2^{NA}&=&-\frac{151}{36}-\frac{13}2\zeta_3
  +\pi^2\left(\frac{179}{72}-\frac83\ln2\right),\nonumber\\
C_2^L&=&\frac{11}9,\qquad C_2^H\ =\ \frac{44}9-\frac49\pi^2.
\label{hcorr}
\end{eqnarray}

\newpage


\end{document}